\documentclass[12pt,a4paper,fleqn]{article}

\usepackage{fullpage}

\usepackage{amsmath}
\usepackage{amssymb}
\usepackage{amscd}
\usepackage{plain}
\usepackage{graphicx}
\usepackage{epsfig}
\usepackage{fancyhdr}
\usepackage{t1enc,epsfig}
\usepackage[mathscr]{eucal}
\usepackage{epstopdf}
\usepackage[german,english]{babel}

\newtheorem{theorem}{Theorem}[section]

\newtheorem{corollary}[theorem]{Corollary}
\newtheorem{definition}[theorem]{Definition}
\newtheorem{example}[theorem]{Example}
\newtheorem{lemma}[theorem]{Lemma}
\newtheorem{proposition}[theorem]{Proposition}
\newtheorem{remark}[theorem]{Remark}
\newtheorem{conjecture}[theorem]{Conjecture}
\numberwithin{equation}{section}

\newenvironment{proof}{{\bf Proof. }}{\hfill$\rule{1ex}{1ex}$\par\medskip}

\newcommand{\R}{\mathbb{R}}

\begin{document}

\newcommand{\bthm}{\begin{theorem}}
\newcommand{\ethm}{\end{theorem}}
\newcommand{\bd}{\begin{definition}}
\newcommand{\ed}{\end{definition}}
\newcommand{\bs}{\begin{proposition}}
\newcommand{\es}{\end{proposition}}
\newcommand{\bp}{\begin{proof}}
\newcommand{\ep}{\end{proof}}
\newcommand{\be}{\begin{equation}}
\newcommand{\ee}{\end{equation}}
\newcommand{\ul}{\underline}
\newcommand{\br}{\begin{remark}}
\newcommand{\er}{\end{remark}}
\newcommand{\bex}{\begin{example}}
\newcommand{\eex}{\end{example}}
\newcommand{\bc}{\begin{corollary}}
\newcommand{\ec}{\end{corollary}}
\newcommand{\bl}{\begin{lemma}}
\newcommand{\el}{\end{lemma}}
\newcommand{\bj}{\begin{conjecture}}
\newcommand{\ej}{\end{conjecture}}

\def\cmp{{\complement}}

\def\tA{{\tt A}}
\def\tB{{\tt B}}
\def\tC{{\tt C}}
\def\tD{{\tt D}}
\def\td{{\tt d}}
\def\tE{{\tt E}}
\def\tte{{\tt e}}
\def\tF{{\tt F}}
\def\tG{{\tt G}}
\def\tg{{\tt g}}
\def\ti{{\tt i}}
\def\tI{{\tt I}}
\def\tj{{\tt j}}
\def\tn{{\tt n}}
\def\tL{{\tt L}}
\def\tO{{\tt O}}
\def\tP{{\tt P}}
\def\tq{{\tt q}}
\def\ttr{{\tt r}}
\def\tP{{\tt P}}
\def\tR{{\tt R}}
\def\tS{{\tt S}}
\def\ttt{\tt t}
\def\tT{{\tt T}}
\def\ttg{{\tt g}}
\def\ttG{{\tt G}}
\def\bttg{\overline{\tg}}
\def\tu{{\tt u}}
\def\tv{{\tt v}}
\def\tV{{\tt V}}
\def\tw{{\tt w}}
\def\tx{{\tt x}}
\def\ty{{\tt y}}
\def\tz{{\tt z}}

\def\bgam{{\mbox{\boldmath$\gamma$}}}
\def\uGam{\underline\Gamma}

\def\boeta{{\mbox{\boldmath$\eta$}}}
\def\oboeta{\overline\boeta}
\def\ups{\upsilon}

\def\Om{\Omega}
\def\om{\omega}
\def\oom{\overline\omega}
\def\bttg{\mbox{\boldmath${\tt g}$}}
\def\btau{\mbox{\boldmath${\tau}$}}
\def\bom{{\mbox{\boldmath$\omega$}}}
\def\obom{\overline\bom}
\def\0bom{{\bom}^0}
\def\0obom{{\obom}^0}
\def\nbom{{\bom}_n}
\def\0nbom{{\bom}_{n,0}}
\def\n*bom{{\bom}^*_{(n)}}
\def\wt{\widetilde}
\def\wtbom{\widetilde\bom}
\def\whbom{\widehat\bom}
\def\oom{\overline\om}
\def\wtom{\widetilde\om}
\def\bOm{\mbox{\boldmath${\Om}$}}
\def\obOm{\overline\bOm}
\def\whbOm{\widehat\bOm}
\def\wtbOm{\widetilde\bOm}

\def\Gam{\Gamma}
\def\Lam{\Lambda}
\def\lam{\lambda}

\def\Ups{\Upsilon}
\def\utheta{\underline\theta}
\def\ovr{\overline r}

\def\oG{\overline G}
\def\oL{\overline L}

\def\bbC{\mathbb C}
\def\bbE{\mathbb E}
\def\bbP{\mathbb P}
\def\fB{\mathfrak B}
\def\fG{\mathfrak G}
\def\fW{\mathfrak W}
\def\bbQ{\mathbb Q}

\def\bi{\mathbf i}
\def\bj{\mathbf j}
\def\bn{\mathbf n}
\def\bt{\mathbf t}
\def\bu{\mathbf u}
\def\bw{\mathbf w}
\def\bX{\mathbf X}
\def\ubX{\underline\bX}
\def\bx{\mathbf x}
\def\ubx{\underline\bx}
\def\bY{\mathbf Y}
\def\by{\mathbf y}
\def\ubY{\underline\bY}
\def\bZ{\mathbf Z}
\def\bz{\mathbf z}

\def\cl{\centerline}

\def\cA{\mathcal A}
\def\cB{\mathcal B}
\def\cC{\mathcal C}
\def\cD{\mathcal D}
\def\cE{\mathcal E}
\def\cF{\mathcal F}
\def\cH{\mathcal H}
\def\cK{\mathcal K}
\def\cL{\mathcal L}
\def\cN{\mathcal N}
\def\cS{\mathcal S}
\def\cT{\mathcal T}
\def\cV{\mathcal V}
\def\cW{\mathcal W}
\def\ocH{\overline\cH}
\def\ocW{\overline\cW}

\def\bbB{\mathbb B}
\def\bbK{\mathbb K}
\def\bbL{\mathbb L}

\def\bbR{\mathbb R}
\def\bbS{\mathbb S}
\def\bbT{\mathbb T}
\def\bbZ{\mathbb Z}
\def\ba{\mathbf a}
\def\bg{\mathbf g}
\def\bX{\mathbf X}
\def\bx{\mathbf x}
\def\wtbx{\widetilde\bx}
\def\ui{{\underline i}}

\def\oA{{\overline A}}
\def\uA{{\underline A}}
\def\ua{{\underline a}}
\def\uua{{\underline{a_{}}}}
\def\oa{{\overline a}}
\def\uk{{\underline k}}
\def\ux{{\underline x}}
\def\wtux{\widetilde\ux}
\def\uX{{\underline X}}
\def\by{\mathbf y}
\def\uy{\underline y}
\def\bY{\mathbf Y}
\def\uY{\underline Y}

\def\uj{{\underline j}}
\def\unn{\underline n}
\def\unp{\underline p}
\def\ovp{\overline p}
\def\bx{\mathbf x}
\def\ox{\overline x}
\def\obx{\overline\bx}
\def\uz{\underline z}
\def\bz{\mathbf z}
\def\uv{\underline v}
\def\dist{\textrm{dist}}
\def\diy{\displaystyle}
\def\ov{\overline}
\def\u0{{\underline 0}}

\def\oomega{\overline\omega}
\def\oUpsilon{\overline\Upsilon}
\def\wtomega{\widetilde\omega}
\def\wtz{\widetilde z}
\def\wtheta{\widetilde\theta}
\def\wtalpha{\widetilde\alpha}
\def\wh{\widehat}
\def\oV{\overline {\mathcal V}}

\def\bI{\mathbf I}
\def\bN{\mathbf N}
\def\bbN{\mathbf N}
\def\bP{\mathbf P}
\def\bV{\mathbf V}
\def\oW{\overline W}
\def\ofW{\overline\fW}
\def\LT{{\mathbb{LT}}}
\def\mucr{{\mu_{cr}}}

\def\rA{{\rm A}}
\def\rB{{\rm B}}
\def\urB{\underline\rB}
\def\rc{{\rm c}}
\def\rC{{\rm C}}
\def\rd{{\rm d}}
\def\rD{{\rm D}}
\def\rd{{\rm d}}
\def\re{{\rm e}}
\def\rE{{\rm E}}
\def\rF{{\rm F}}
\def\rI{{\rm I}}

\def\rn{{\rm n}}

\def\rO{{\rm O}}
\def\rP{{\rm P}}
\def\rQ{{\rm Q}}
\def\rr{{\rm r}}
\def\rR{{\rm R}}

\def\rs{{\rm s}}
\def\rS{{\rm S}}
\def\rT{{\rm T}}
\def\rV{{\rm V}}

\def\rw{{\rm w}}

\def\rx{{\rm x}}
\def\ry{{\rm y}}
\def\rtr{\rm{tr}}

\def\oa{\overline a}
\def\ua{\underline a}

\def\uk{\underline k}
\def\un{\underline n}
\def\ux{\underline x}
\def\uy{\underline y}
\def\wtux{\widetilde\ux}
\def\uX{\underline X}

\def\oJ{\overline J}
\def\oP{\overline P}
\def\utC{{\underline\tC}}
\def\utD{{\underline\tD}}
\def\utE{{\underline\tE}}
\def\urB{{\underline\rB}}
\def\urC{{\underline\rC}}
\def\urD{{\underline\rD}}
\def\urE{{\underline\rE}}

\def\vng{{\varnothing}}
\def\cl{\centerline}

\def\BX{{\mathbf X}} \def\bx{{\mathbf x}}
\def\bbZ{{\mathbb Z}} \def\bbP{{\mathbb P}}
\def\bz{\mathbf z}

\def\cB{{\mathcal B}} \def\cX{{\mathcal X}}
\def\fB{\mathfrak B}\def\fM{\mathfrak M} \def\fX{\mathfrak X}
\def\cT{\mathcal T}
\def\bu{\mathbf u}
\def\bv{\mathbf v}\def\bx{\mathbf x}\def\by{\mathbf y}
\def\om{\omega} \def\Om{\Omega}
\def\bbP{\mathbb P} \def\hw{{h^{\rm w}}} \def\hwphi{{h^{\rm w}_\phi}}
\def\beq{\begin{eqnarray}} \def\eeq{\end{eqnarray}}
\def\beqq{\begin{eqnarray*}} \def\eeqq{\end{eqnarray*}}

\def\rb{{\rm b}}
\def\rd{{\rm d}} \def\rmv{{\rm v}} \def\rV{{\rm V}}
\def\Dwphi{{D^{\rm w}_\phi}}
\def\BX{\mathbf{X}}
\def\hwphiii{{h^{\rm w}_{\phi_1\otimes\phi_2\otimes\dots\otimes \phi_n}}}
\def\hwphii{{h^{\rm w}_{\phi_1\otimes\phi_2}}}
\def\mwe{{D^{\rm w}_\phi}}
\def\DwPhi{{D^{\rm w}_\Phi}} \def\iw{i^{\rm w}_{\phi}}
\def\bE{\mathbb{E}} \def\mbE{\mathbf{E}}
\def\mbp{{\mathbf p}}
\def\1{{\mathbf 1}} \def\fB{{\mathfrak B}}  \def\fM{{\mathfrak M}}
\def\bbE{{\mathbb E}}

\def\tha{{\theta}} \def\uBX{{\underline\BX}}
\def\gam{\gamma} \def\kap{\kappa} \def\lam{\lambda}
\def\ups{{\upsilon}}

\def\vphi{\varphi} \def\vpi{\varpi}
\def\veps{\varepsilon} \def\vrho{\varrho}

\def\blam{{\mbox{\boldmath${\lambda}$}}} \def\bphi{{\mbox{\boldmath${\phi}$}}}
\def\bpsi{{\mbox{\boldmath${\psi}$}}} \def\bta{{\mbox{\boldmath${\eta}$}}}
\def\bzeta{{\mbox{\boldmath${\zeta}$}}} \def\btau{{\mbox{\boldmath${\tau}$}}}
\def\bups{{\mbox{\boldmath${\ups}$}}}
\def\bUps{{\mbox{\boldmath${\Ups}$}}}
\def\bu{\mathbf u}
\def\bU{\mathbf U}
\def\bT{\mathbf T}
\def\bpi{{\mbox{\boldmath${\pi}$}}}

\def\lam{{\lambda}} \def\eps{{\epsilon}}
\def\Lam{{\Lambda}} \def\BK{{\mathbf K}} \def\BP{{\mathbf P}}
\def\bOne{\mathbf 1} \def\uOne{\underline 1} \def\ulam{{\underline\lam}}
\def\bbT{\mathbb T}

\def\urM{{\underline{\rm M}}}
\def\uPi{{\underline\Pi}}
\def\rC{{\rm C}} \def\rL{{\rm L}}  \def\rM{{\rm M}}
\def\rP{{\rm P}} \def\rR{{\rm R}} \def\cA{\mathcal A}
 \def\cB{\mathcal B} \def\cF{\mathcal F} \def\cG{\mathcal G}
\def\cM{\mathcal M} \def\cV{\mathcal V}
\def\cX{\mathcal X} \def\cY{\mathcal Y}

\def\rt{{\rm t}} \def\ru{{\rm u}}
\def\rv{{\rm v}} \def\rw{{\rm w}}

\def\bbR{{\mathbb R}}
\def\bbA{{\mathbb A}} \def\bbB{{\mathbb B}}
\def\bbC{{\mathbb C}}  \def\bbE{{\mathbb E}}
\def\bbD{{\mathbb D}}\def\bbG{{\mathbb G}}
\def\bbM{{\mathbb M}} \def\bbP{\mathbb P}
 \def\bbS{{\mathbb S}}
  \def\bbZ{{\mathbb Z}}

\def\ree{{\rm e}}

\def\tA{{\tt A}} \def\tB{{\tt B}} \def\tC{{\tt C}} \def\tI{{\tt I}}
\def\tJ{{\tt J}} \def\tK{{\tt K}}
\def\tL{{\tt L}} \def\tP{{\tt P}} \def\tQ{{\tt Q}}
\def\tR{{\tt R}} \def\tS{{\tt S}}
\def\tW{{\tt W}}
\def\ty{{\tt y}} \def\tz{{\tt z}}
\def\t0{{\tt 0}} \def\tp{{\tt p}}

\def\fy{{\mathfrak y}} \def\fz{{\mathfrak z}}

\def\iGam{{\mathit{\Gam}}}
\def\igam{{\mathit{\gamma}}}
 \def\iUps{{\mathit{\Ups}}} \def\ups{{\upsilon}}
\def\iups{{\mathit{\ups}}}
\def\itau{{\mathit{\tau}}}

\def\be{\begin{equation}}
\def\ee{\end{equation}}

\def\beal{\begin{array}{l}}
\def\beac{\begin{array}{c}}
\def\bear{\begin{array}{r}}
\def\beacl{\begin{array}{cl}}
\def\beacr{\begin{array}{cr}}
\def\ena{\end{array}}

\def\diy{\displaystyle}

\def\sA{\mathscr A} \def\sB{\mathscr B} \def\sC{\mathscr C}
\def\sF{\mathcal F} \def\sG{\mathcal G} \def\sI{\mathcal I}
\def\sL{\mathcal L}\def\sM{\mathscr M} \def\sO{\mathcal O}
\def\sP{\mathcal P} \def\sR{\mathcal R} \def\sS{\mathcal S}

\title{On principles of large deviation and selected data compression}

\author{Y. Suhov$^1$, I. Stuhl$^2$}

\date{}
\footnotetext{

\noindent $^1$ Mathematics Dept., Penn State University, University Park, State College,
PA 16802, USA; DPMMS, University of Cambridge, UK;\\
E-mail: yms@statslab.cam.ac.uk

\noindent $^2$ Mathematics Dept.,
University of Denver, Denver, CO 80208 USA;
 Appl. Math and Prob. Theory Dept., University of Debrecen, Debrecen, 4028, HUN;\\
E-mail: izabella.stuhl@du.edu}

\maketitle

\begin{abstract} The Shannon Noiseless coding theorem (the data-compression
principle) asserts that for an information source with an alphabet $\cX=\{0,\ldots ,\ell -1\}$
and an asymptotic equipartition property, one can reduce the number
of stored strings $(x_0,\ldots ,x_{n-1})\in\cX^n$ to
$\ell^{nh}$ with an arbitrary small error-probability. Here $h$ is
the entropy rate of the source (calculated to the base $\ell$). We
consider further reduction based on the concept
of utility of a string measured in terms of a rate of a weight
function. The novelty of the work is that the distribution of memory
is analyzed from a probabilistic point of view. A convenient tool
for assessing the degree of reduction is a probabilistic large deviation
principle. Assuming a Markov-type setting, we discuss some relevant
formulas, including the case of a general alphabet.
\end{abstract}

\section{Introduction}

Consider a discrete-time random
process $\BX =(X_n)$, $n\in\bbZ_+:=\{0,1,2,\ldots\}$, where the random
variable $X_n$ -- possibly a random vector or a random element in a space
$\cX$ -- describes the state of
the process at time $n$. One interpretation used throughout the paper is that
process $\BX$ represents an {\it information source}, in the spirit of
\cite{CT}, \cite{KS};
here set $\cX$ will play role of a source alphabet. Under
such an interpretation the probability distribution of $\BX$ (on
$\cX^{\bbZ_+}$) is referred to as $\bbP^{\rm{so}}$.  Sample states
of the process are given by points $x\in\cX$.  An (initial) $n$-{\it string}
is a collection $\bx_0^{n-1}=\{x_i:\,0\leq i<n\}\in\cX^n$; $n$ is referred
to as the length of $\bx_0^{n-1}$. A random sample drawn from $\BX$ is
denoted by $\BX_0^{n-1}$; it is a random element in $\cX^n$. The probability
distribution for $\BX_0^{n-1}$ generated by $\bbP^{\rm{so}}$ is denoted by
$p^{\rm{so}}_n$ (i.e., $\bbP^{\rm{so}}(\BX_0^{n-1}\in\cB_n)=p^{\rm{so}}_n
(\cB_n)$, for any (Borel) set $\cB_n\subseteq\cX^n$). For a process with
discrete states (with a finite or countable
alphabet $\cX$), the value $p^{\rm{so}}_n(\bx_0^{n-1})=\bbP^{\rm{so}}
(\BX_0^{n-1}=\bx_0^{n-1})$. In this context, the concepts of information
and entropy rates are relevant; see below.

However, there are situations where one may need to extend (or complement)
standard notions. In this work we are motivated by Refs \cite{SSYK, SYS}
discussing {\it weighted} information and entropy. These concepts emerge
when one introduces a {\it weight function} $\phi_n(\bx_0^{n-1})$
reflecting {\it utility} of an outcome string $\bx_0^{n-1}$.

A second interpretation emerges when we consider the
problem of {\it storing} strings $\bx_0^{n-1}$. Suppose we have a notion
of `volume' in $\cX$ associated with a measure $\nu$ with $\rV =\nu (\cX
)<\infty$ (e.g., the number of points in a set $\cA\subseteq\cX$ in the
case of a finite alphabet). Then the volume in $\cX^n$ may be represented
by the product-measure $\nu^n$. A normalized volume $\diy\frac{\nu^n(\cB_n)}
{\rV^n}$, $\cB_n\subseteq\cX^n$, gives a probability distribution on
$\cX^n$ (with IID digits), and an (asymptotic) analysis of $\nu^n$ is
reduced to an analysis of this probability distribution. When the cardinality
$\#\,(\cX)=\ell$ is finite and $\nu (\cA )=\#\,\cA$ (a counting measure),
we obtain $\rV =\ell$. The volume of a set $\cB_n\subseteq\cX^n$ is written
as $\#\,\cB_n=\ell^n p^{\rm{eq}}_n(\cB_n)$ where  $p^{\rm{eq}}_n$ stands
for an equidistribution on $\cX^n$, with
$p^{\rm{eq}}_n(\bx_0^{n-1})=1/\ell^n$ for all $\bx_0^{n-1}\in\cX^n$.

More generally, we can think of a probability distribution $p^{\rm{st}}_n$
on $\cX^n$ such that the volume in $\cX^n$ is represented by
$\rV_np^{\rm{st}}_n(\cB_n)$, $\cB_n\subseteq\cX^n$, where $\rV_n$ is a given
constant (yielding the total amount of memory (or space in a broader sense)
available for storing strings of length $n$). Then
asymptotic properties of $p^{\rm{st}}_n$ can be used for
assessing the volume of random strings $\BX_0^{n-1}$ generated by
$p^{\rm{so}}_n$. In this paper, such an
approach is used for the purpose of {\it selected data compression}.

Returning to the information source interpretation, the standard
(Shannon) information $I(\bx_0^{n-1})$ and entropy $H(p^{\rm{so}}_n)$ of
the source $n$-string is given by
\be\label{eq:InfEnt}\beac
I(\bx_0^{n-1})= -\log\,p^{\rm{so}}_n(\bx_0^{n-1}), \qquad
H(p^{\rm{so}}_n) = \sum\limits_{\bx_0^{n-1}\in\cX^n}p^{\rm{so}}_n(\bx_0^{n-1})
I (\bx_0^{n-1}).\ena\ee
The rates
\be\label{eq:eirates}i=\lim\limits_{n\to\infty}\diy
\frac{I(\BX_0^{n-1})}{n}\;\hbox{ $\bbP^{\rm{so}}$-a.s.,\;\, and \,\;}
h=\lim\limits_{n\to\infty}\diy\frac{H(p^{\rm{so}}_n)}{n}\ee
are fundamental parameters of a random process leading to profound results
and fruitful theories with far-reaching consequences, cf. \cite{CT, KS}.$^{1)}$
\footnote{$^{1)}$As a rule (with exceptions), references
of a general character are given to books rather than to original
papers.}In fact, under mild assumptions, $h=i$: this is the
Shannon--McMillan--Breiman theorem
\cite{CT,AC}.

In this paper, we treat two types of weight functions $\phi_n(\bx_0^{n-1})$:
additive and multiplicative; see below. A justification of our approach can
be provided through aforementioned selected data compression. The basic
idea of the Shannon Noiseless coding theorem (NCT), or data-compression (DC),
was to disregard strings/messages $\bx_0^{n-1}$ of length $n>>1$ (drawn from
$p^{\rm{so}}_n$) which are highly unlikely. (That is, with low probabilities
$p^{\rm{so}}_n(\bx_0^{n-1})$, or, equivalently, with high information $I(\bx_0^{n-1})$,
for discrete outcomes.) Incidentally, one also
disregards strings that are highly likely. The remaining strings, forming
set $\cT_n\subset\cX^n$ with $p^{\rm{so}}_n(\cT_n)\to 1$, can be
characterized through the information/entropy rate (IER) $h=i$ by invoking the asymptotic
equipartition property (AEP). Pictorially, all strings $\bx_0^{n-1}\in\cT_n$
carry, approximately, the same IER $i=h$; cf.
\eqref{eq:eirates}. Assume, until a further note, that the total number
of $n$-strings equals $\ell^n$ where $\#\,\cX =\ell <\infty$. Then the DC
allows us to diminish the amount of memory
needed to store strings $\bx_0^{n-1}\in\cT_n$ by reducing their length
from $n$ to $\diy\frac{nh}{\log\,\ell}$. (Such a reduction is effectuated
by a lossless coding.) Here $h\leq \log\,\ell$ (and in many realistic
situations, $h<< \log\,\ell$). The
probability $p^{\rm{so}}_n(\cX^n\setminus\cT_n)$ of information loss is
kept small because
of the AEP (which is a Law of large numbers for $I(\BX_0^{n-1})$.)

Now, one may be interested in a further reduction
of the used memory by extracting and storing only `valuable' strings.
It can be done by using a given weight function (WF)
$\phi_n$: strings  $\bx_0^{n-1}$ with high growth rates of
$\phi_n(\bx_0^{n-1})$ are stored while others disregarded.

We show that selecting most valued strings yields a
further reduction of the storage memory, and its effect can
be estimated numerically. The number of strings
in a selected set $\cB_n\subseteq\cT_n$ is given by
$\#\,\cB_n=\ell^np^{\rm{eq}}_n(\cB_n)$. Hence,
\be \beac\label{eq:gamkap}\lim\limits_{n\to\infty}\diy{\frac{1}{n}}
\log\,\#\,\cB_n=\gam\;\hbox{ where }\;\gam:=\log\ell+\kap, \qquad
\kap:=\lim\limits_{n\to\infty}{\diy\frac{1}{n}}\log\,p^{\rm{eq}}_n(\cB_n).\ena \ee
More generally, when we measure the volume of $\cB_n$ by
$\rV_np^{\rm{st}}_n(\cB_n)$, we encounter the limit $\gam =v+\kap$. Here
$v=\lim\limits_{n\to\infty}\;{\diy\frac{1}{n}}\log\rV_n$ and
\be\label{eq:kapgg}\kap:=\lim\limits_{n\to\infty}\frac{1}{n}
\log\,p^{\rm{st}}_n(\cB_n).\ee
Passing from $p^{\rm{eq}}_n$ to $p^{\rm{st}}_n$ makes the amount of memory
needed to store $\bx_0^{n-1}$ string-dependent and separate the issues of
the total volume $\rV_n$ and that of the distribution of the storage
volume between different strings.

The value $\kap$ in \eqref{eq:kapgg} can be studied via the Large deviation (LD)
theory. The LD studies are now an established trend in theoretical and applied probability;
some important reference sources are  \cite{DZ}--\cite{V2}. The core is the LD principle (LDP);
its gist (as we use it in this work) is summarized in the formula
\be\label{eq:kapinco}\kap =-\inf\;\Big[\Pi^*(z):\;z \in B\Big].\ee
Here $B$ is a set of probabilistic vectors (empirical measures) in a
(suitable) Euclidean space ($\bbR^\ell$ or $\bbR^{\ell^2}$
and so on); the form of $B$ depends upon the choice of sets $\cB_n$. (Set $B$ is
constructed from a frequency analysis of strings $\bx_0^{n-1}\in\cB_n$
and turns out to be a convex polyhedron.) Next,
$\Pi^*$ is a large deviation rate (LDR) function. Typically, $\Pi^*$ is
a lower semi-continuous convex function representing the Legendre--Fenchel
transform of some
moment-generating function $\Pi$ (this fact is encapsulated in
the G\"artner--Ellis theorem). Furthermore, $\Pi^*$
has a form of a relative entropy (an observation going back to the 1957
Sanov theorem; cf. e.g., \cite{CT} and \cite{DZ}--\cite{V2}). Next, $\Pi^*(z)=0$
at a point (or points) $z$ representing  a related expected value and $\Pi^*(z)>0$
at all other points $z$ (this includes values $\Pi^*(z)=+\infty$).
It yields that $\gam \leq\log\ell$,
and in many cases $\gam<<\log\ell$, depending on the choice of $\cB_n$.
Consequently, the reduction in length is from $n$ to $\diy\frac{n\gam}{\log\ell}$.
(In our situation, as $\cB_n\subset\cT_n$, value $\gam$ will be $<h$, achieving
a distinct improvement compared with the Shannon NCT.)
It is helpful that both $B$ and $\Pi^*$ admit
some `standard' representations reflecting the structure of $p^{\rm{so}}_n$
(and $\phi_n$) in a natural
(and computationally convenient) manner allowing to calculate the value
$\gam$. This is the thrust of our approach: the LDP is used for
$p^{\rm{st}}_n$ whereas $p^{\rm{so}}_n$ and $\phi_n$ specify $B$ and
$\Pi^*$ (through selected sets $\cB_n$).

A basic condition adopted in this paper is that each of distributions
$p^{\rm{so}}_n$ and $p^{\rm{st}}_n$ is generated by a discrete-time Markov
chain (DTMC). Although the LDP scheme is formally applicable in a more
general situation, the Markov assumption will allow us to simplify
technicalities.
For additive and multiplicative WFs $\phi_n$
we write down formulas for the value $\kap$ in \eqref{eq:kapgg}
and \eqref{eq:kapinco} and specify them when $p^{\rm{st}}_n=p^{\rm{eq}}_n$.
As was said above, such an approach allows for varying both the set $\cB_n$
(that is, threshold values for utility and information carried by
a selected string) and the
distribution $p^{\rm{st}}_n$ of the normalized volume allocated to
different strings. In practical
terms, it means that one can predict an impact of an adaptation of storage
principles to changing demands and conditions.

In Sect \ref{sect:finite} we
deal with the case of a finite alphabet with $\#\,\cX =\ell$, cf.
\eqref{eq:kapPB0}, \eqref{eq:kapeq0}, \eqref{eq:gameq0}, \eqref{eq:gamIID0};
Sect \ref{sect:contin} treats a general case, cf. \eqref{eq:kap11}, \eqref{eq:gameqg},
\eqref{eq:gamIIDg}. 
As was said, we focus upon two kinds of WFs: (a) additive and
(b) multiplicative. In the simplest form:
\be\label{eq:WFsAM} {\rm{(a)}}\;\phi_n(\bx_0^{n-1})
=\sum\limits_{i=0}^{n-1}\vphi_1(x_i),\qquad
{\rm{(b)}}\;\phi_n(\bx_0^{n-1})=\prod\limits_{i=0}^{n-1}\psi_1(x_i). \ee
Here $x\in\cX\mapsto\vphi_1(x)$ and $x\in\cX\mapsto\psi_1(x)$ are given
functions (one-digit WFs); for brevity, we write $\vphi (x)$ and $\psi (x)$.
Additive WFs may emerge in relatively stable situations where each
observed digit $X_i$
brings reward or loss $\vphi (X_i)$; the utility value $\phi_n(\BX_0^{n-1})$
is treated as a cumulative gain or
deficit after $n$ trials. Multiplicative WFs reflect a more turbulent
scenario where the value (e.g., a fortune; see \cite{SSK})
increases/decreases by a factor $\psi (X_i)$ when outcome $X_i$ is observed.

The topic of this work is closely related to the topic of weighted information/weighted
entropy rates; see \cite{SS1}.

\vskip 3 truemm

{\bf N.B.} In this work we do {\bf not} claim new LD results, offering instead some new
prospects of the LD methodology.

\section{Selected data-compression for a finite-alphabet Markov source}\label{sect:finite}

\subsection{}
We start with additive one-digit WFs $\phi_n(\bx)=\sum\vphi (x_i)$,
$\bx =(x_0,\ldots ,x_{n-1})$; cf. \eqref{eq:WFsAM}.
Assume that probabilities $p^{\rm{st}}_n$ are generated by an irreducible
and aperiodic DTMC where the state space $\cX=\{0,\ldots ,\ell-1\}$. 
Let $\tP^{\rm{st}}=(\tp^{\rm{st}}_{ij})$ and $\lam =(\lam (j))$ designate
the transition matrix (TM) and an initial distribution. Then
$p^{\rm{st}}_n(\bx )=\lam (x_0)\prod\tp^{\rm{st}}_{x_ix_{i+1}}$.
We will analyze occupancy-fraction vectors
$\bU^{(n)}=(U^{(n)}_i)$ and $\bT^{(n)}=(T^{(n)}_{ij})$, of dimensions $\ell$
and $\ell^2$. For $i,j\in\cX$, entries $U^{(n)}_j=U^{(n)}_j(\bx )$ and
$T^{(n)}_{j,j'}=T^{(n)}_{j,j'}(\bx )$ are given by
\be\label{eq:UT}\beal U^{(n)}_i={\diy\frac{1}{n}}\sum\limits_{l=0}^{n-1}
{\mathbf 1}_{x_l=i},\qquad\qquad T^{(n)}_{i,j}={\diy\frac{1}{n-1}}\sum\limits_{l=0}^{n-2}
{\mathbf 1}_{x_l=i,x_{l+1}=j},\\
\hbox{with $\phi_n(\bx)=n\sum\limits_{i\in\cX}^\ell U^{(n)}_i\vphi (i)$ \,\,and}\\
\hbox{$\log\,p^{\rm{st}}_n(\bx)=\log\lam (x_0)+
(n-1)\sum\limits_{i,j\in\cX}^\ell T^{(n)}_{ij}\log\,\tp^{\rm{st}}_{ij}$.}\ena\ee

Assume in addition that the source probabilities $p^{\rm{so}}_n$ are also
generated by an irreducible and aperiodic DTMC with a TM
$\tP^{\rm{so}}=(\tp^{\rm{so}}_{ij})$ and equilibrium distribution
$\pi^{\rm{so}}=(\pi^{\rm{so}}_i)$. The IER $h$  in \eqref{eq:eirates}
takes the form $h=-\sum\limits_{i,j}\pi^{\rm{so}}_i\tp^{\rm{so}}_{ij}
\log\,\tp^{\rm{so}}_{ij}$.

Given $\eps, \eta >0$, the selected set $\cB_n =\cB_n (\eps ,\eta )\subseteq\cX^n$
is
\be\label{eq:setBn0} \cB_n=\Big\{\bx :\hbox{$\sum\limits_{i}
U^{(n)}_i\vphi (i)\geq\eta$,}\,\,\hbox{$-\sum\limits_{i,j}
T^{(n)}_{ij}\log\tp^{\rm{so}}_{ij}\leq h+\eps$}\Big\}.\ee
We choose $\cB_n$ to be a subset in
$\cT_n=\{\bx :\;T^{(n)}_{ij}\log\tp^{\rm{so}}_{ij}\leq h+\eps\}$,
the set which has probability $p^{\rm{so}}_n(\cT_n)\to 1$ but it
is not required by our method.

We will use the LDP under $p^{\rm{st}}_n$ (in our case for vectors
$\bU^{(n)}$ and $\bT^{(n)}$)
to assess the volume of $\cB_n$. The corresponding LDR functions are
denoted by $\rM^*(y)$ and $\Pi^*(z)$; they are specified below, in
\eqref{eq:M*f1Pi*g0}.
(Function $\Pi^*(z)$ can be considered as a natural `extension' of
$\rM^*(y)$.) The form of the LDP is standard, and we do not write
it in detail for the sake of  economy of space.

\subsection{}
Consider the $(\ell -1)$- and $(\ell^2 -1)$-dimensional
{\it simplexes} of probability vectors $y =(y_i)\in\R^\ell$ and
$z =(z_{ij})\in\R^{\ell^2}$, respectively:
$$\beac {\bbS}_\ell=\left\{y =(y_i):\;y_i\geq 0,\;
\sum\;y_i=1\right\},\\
{\bbS}_{\ell^2}=\left\{z =(z_{ij}):\;z_{ij}\geq 0,\;
\sum\limits\; z_{ij}=1\right\}.\ena$$
Then
$\rM^*(y)=\infty$ for $y\in\bbR^\ell\setminus\bbS_\ell$ and
$\Pi^*(z)=\infty$ for $z\in\bbR^{\ell^2}\setminus\bbS_{\ell^2}$.
Given $y =(y_j)\in\bbS_\ell$, $z =(z_{ij})\in\bbS_{\ell^2}$ and
$u =(u_l)\in\bbS_\ell$, set:
\be\label{eq:M*f1Pi*g0}\beac\rM^*(y )=\operatornamewithlimits{\sup}\limits_{
u\in{\bbS}_\ell}\Big[\sum\limits_{j} y_j
\log\,{\diy\frac{u_j}{(\tP^{\rm{st}}u)_j}}\Big]\\
\Pi^*(z )\;=\;\operatornamewithlimits{\sup}\limits_{u\in{\bbS}_\ell}
\Big[\sum\limits_{i,j} z_{ij}
\log\,{\diy\frac{u_j}{(\tP^{\rm{st}}u)_j}}\Big]\ena
\quad\hbox{where }\;(\tP^{\rm{st}}u)_j=\sum\limits_l \tp^{\rm{st}}_{jl}u_l.\ee
Functions
$\rM^*$ and $\Pi^*$ are determined by TM
$\tP^{\rm{st}}$: $\rM^*(y )=\rM^*(\tP^{\rm{st}};y )$ and
$\Pi^*(z)=\Pi^*(\tP^{\rm{st}};z)$. This is a standard form of
an LDR function for occupancies in a Markov case (which holds in
a  more general situation). Cf.
\cite{DS}, Ch. 4.1, particularly Lemmas 4.1.36 and 4.1.40, Theorem 4.1.43
and Lemma 4.1.45, and \cite{DZ}, Ch. 6.5, especially Theorems 6.5.2 and 6.5.4.
A simple explicit formula for $\rM^*(y)$ when $\ell =2$
was proposed in \cite{DM}. See also Sect 2.4 below.

Define a convex polyhedron $B=B(\tP^{\rm{so}},\eps ,\eta )
\subseteq\bbS_{\ell^2}$:
\be\label{eq:setB0}\beal B=\bigg\{z :
-\sum\limits_{i,j}z_{ij}\log\,\tp^{\rm{so}}_{ij}\leq h+\eps,\;
\sum\limits_{i,j}z_{ij}\vphi (i)\geq\eta\;\bigg\}.\ena\ee
Applying general LD results yields

\bthm \label{Thm:2.1}
For all $\eps ,\eta >0$ and initial
distribution $\lam$, the following relation holds for
$\kap =\kap(\tP^{\rm{eq}},\tP^{\rm{so}},\eps,\eta)$:
\be\label{eq:kapPB0}\beac \kap
:=\lim\limits_{n\to\infty}{\diy\frac{1}{n}}\log\,p^{\rm{st}}_n(\cB_n)
=-\inf\;\Big[\Pi^*(z ):\;z \in B\Big].\ena\ee
Here $\cB_n$, $\Pi^*$ and $B$ are as in \eqref{eq:setBn0}
-- \eqref{eq:setB0}.
\ethm

Further, suppose TM $\tP^{\rm{st}}$ has entries of the form
$\tp^{\rm{st}}_{ij}=\tp_j$
where vector $\mbp =(\tp_j)\in{\bbS}_\ell$. Then $\tP^{\rm{st}}u=\mbp$
for all $u\in{\bbS}_\ell$, and \eqref{eq:M*f1Pi*g0} for $\rM^*(y)$
features the relative entropy
$D(y ||\mbp )=\sum y_j\log\,\diy\frac{y_j}{\tp_j}$. Namely,
\be\label{eq:M*IID}\beac \rM^*(y )=\sup\;\left[\sum y_j
\log\,{\diy\frac{u_j}{\tp_j}}:\;u=(u_j)\in{\bbS}_\ell\right]
=\sum y_j \log\,{\diy\frac{y_j}{\tp_j}}\ena \ee
whenever $y =(y_j)\in{\bbS}_\ell$, in agreement with the Sanov theorem. Thus,
the value $\rM^*(y )$ in \eqref{eq:M*f1Pi*g0} can
be considered as an analog of relative entropy $D(y||\mbp)$ where vector
$\mbp$ is replaced by $\tP^{\rm{st}}$, a stochastic TM.

For $p^{\rm{st}}_n=p^{\rm{eq}}_n$, vector $\mbp =\mbp^{\rm{eq}}
=(1/\ell ,\ldots ,1/\ell )$, and
\be\label{eq:M*H}\beal\rM^*(y )=D(y ||\mbp^{\rm{eq}})=\log\,\ell -H(y )\;
\hbox{where}\\\;\\
\hbox{$H(y )=-\sum  y_i\log\,y_i$ \,\,is the entropy of \,\,$y=(y_j)$.}
\ena\ee
In this case, \eqref{eq:kapPB0} yields (see also \eqref{eq:gamkap}):
\be\label{eq:kapeq0}\kap =\kap (\tP^{\rm{eq}},\tP^{\rm{so}},\eps ,\eta )
=-\log\ell +\gam .\ee
Here $\gam =\gam (\tP^{\rm{so}},\eps,\eta )$ is a supremum on the set
$A=A(\tP^{\rm{so}},\eps,\eta )\subset\bbS_\ell$:
\be\label{eq:gameq0} \gam =\sup\;\Big[H (y):\;
y \in A\Big]\ee
where
\be\label{eqsetAg}\beal A=\Big\{y=(y_i):\;\sum\limits
y_i\vphi (i)\geq\eta\; \hbox{ and $\;\exists\;$ a vector } \,z =(z_{ij})\in\bbR^\ell_+ \\ \;\\
\qquad\hbox{ with } \sum\limits_j z_{ij}=y_i\;\;\forall\;\;i\in\cX
\,\,\hbox{and }\, - \sum\limits_{i,j}z_{ij}\log\,\tp^{\rm{so}}_{ij} \; \leq \;h+\eps\;\Big\} \ena \ee
and $H(y )$ is as in \eqref{eq:M*H}. Again observe that $A\subseteq\bbS_\ell$
is a convex polyhedron. Since $y\in\bbS_\ell\mapsto H(y )$ is a (strictly)
concave function, we have a dichotomy.
Either point $(1/\ell ,\ldots ,1/\ell)\in A$ in which case
$\gam (\tP^{\rm{so}},\eps,\eta )=\log\,\ell$
or else $(1/\ell ,\ldots ,1/\ell)\not\in A$, $\gam (\tP^{\rm{so}},\eps,\eta )
<\log\,\ell$, and the supremum in \eqref{eq:kapeq0}
attained at a single point in the boundary $\partial A$ reached by the
corresponding level surface of $H(y)$.

If TM $\tP^{\rm{so}}$ has $\tp^{\rm{so}}_{ij}=\tp^{\rm{so}}_j$ (an IID source),
\eqref{eq:gameq0} is simplified. Introduce
vector $\mbp^{\rm{so}} =(\tp^{\rm{so}}_j)$: here the IER
$h=-\sum\limits_i \tp^{\rm{so}}_i\log\,\tp^{\rm{so}}_i$, and
$\gam =\gam (\mbp^{\rm{so}},\eps ,\eta )$ is given by
\be\label{eq:gamIID0} \gam =\sup\;\Big[H (y):\;
y \in D\Big],\ee
where polyhedron $D=D(\mbp^{\rm{so}},\eps,\eta )\subset\bbS_\ell$:
\be\label{eq:Dpoly} D=\Big\{y=(y_j)\in{\bbS}_\ell:\;\sum y_i\vphi (i)\geq\eta
\quad\hbox{and}\quad-\sum\limits y_i\log\,\tp^{\rm{so}}_i
\leq h+\eps\Big\}.\ee

\noindent
Let us summarize. For an additive WF $\phi_n(\bx)
=\sum\vphi (x_i)$ the following result emerges:

\bthm \label{Thm:2.2}
Assume the source probabilities $p^{\rm{so}}_n$
are generated by an irreducible and aperiodic, stationary DTMC
with an alphabet $\cX=\{0,\ldots ,\ell -1\}$, TM $\tP^{\rm{so}}=(\tp^{\rm{so}}_{ij})$
and equilibrium distribution $\pi^{\rm{so}}=(\pi^{\rm{so}}_i)$. Set
$h=-\sum\limits_{i,j}\pi^{\rm{so}}_i\tp^{\rm{so}}_{ij}
\log\,\tp^{\rm{so}}_{ij}$. When selecting strings
$\bx\in\cX^n$ with
\be\label{eq:cBn0}{\diy\frac{1}{n}}\sum\limits_{i=0}^{n-1}\vphi (x_i)
\geq \eta\hbox{ and }-\frac{1}{n-1}
\log\,p^{\rm{so}}_n(\bx )\leq h+\eps ,\ee
the number $b_n$ of selected strings satisfies
\be\label{eq:cBn01}\lim\limits_{n\to\infty}\frac{1}{n}\log\,b_n=
\gam (\tP^{\rm{so}},\eps,\eta )\ee
where $\gam (\tP^{\rm{so}},\eps,\eta )$ is given by \eqref{eq:kapeq0}.
For an IID source, with $\tp^{\rm{so}}_{ij}=\tp^{\rm{so}}_j$, one uses
\eqref{eq:gamIID0} with $h=
-\sum\limits_{j=1}^\ell\tp^{\rm{so}}_j\log\,\tp^{\rm{so}}_j$.
\ethm

\subsection{}
For completeness, we state an assertion for an
WF $\phi_n (\bx_0^{n-1})=\sum\limits_{i=0}^{n-k}
\vphi (\bx_{i}^{i+k-1})$ (when the summand WF $\vphi$ takes into account
$k$ previous digits produced by the source) where $\bx_{i}^{i+k-1}
=(x_i,\ldots ,x_{i+k-1})$. Here we select strings
$\bx_0^{n-1}\in\cX^n$ with
\be\label{eq:cBn2}{\diy\frac{1}{n-k}}\sum\limits_{i=0}^{n-k}\vphi (\bx_i^{i+k-1})\geq \eta ,\qquad
\frac{-1}{n-k}\log\,p^{\rm{so}}_n(\bx_0^{n-1})\leq h+\eps \ee
where $h$ is as in \eqref{eq:eirates}. With
$\cX =\{0,\ldots ,\ell -1\}$,
assume that $p^{\rm{so}}_n$ are generated by a DTMC of order $k$,
with state space $\cX^k$, $k$-step transition probabilities
$\tp^{\rm{so}}_{\bu,\bu'}$, $\bu,\bu'\in\cX^k$,
irreducible and aperiodic. Let $\pi^{\rm{so}}_\bu$ stand for the
equilibrium probabilities and
set
\newline $h=-{\diy\frac{1}{k}}\sum\limits_{\bu,\bu'\in\cX^k}\pi^{\rm{so}}_\bu
\tp^{\rm{so}}_{\bu,\bu'}
\log\,\tp^{\rm{so}}_{\bu,\bu'}$.

\bthm Adopt the above assumption. Similarly to \eqref{eq:cBn0},
\eqref{eq:cBn01}, the number $b_n$ of selected strings satisfies
\be\label{eq:cBgn}\lim\limits_{n\to\infty}\frac{1}{n}\log\,b_n=
\gam =\gam (\eps,\eta ).\ee
Here $\gam$ is as follows: for $y=(y_i)\in\bbS_\ell$, set
 $H(y )=-\sum\limits_{i\in\cX}y_i\log\,y_i$, as in \eqref{eq:M*H}. Then
\be\label{eq:gamGen} \gam =\inf\;\Big[H (y):\;
y \in B_{\ell ,k} (\eps,\eta )\Big],\ee
with $B_{\ell ,k}=B_{\ell ,k} (\eps,\eta )\subset\bbS_\ell$:
\be\label{eq:Blkset}\beal B_{\ell ,k}=\bigg\{y=(y_j):\;\hbox{$\exists$ a map}\,
\,\bu =(u_1,\ldots ,u_k)\in\cX^k\mapsto \zeta (\bu )\geq 0\; \hbox{such that}\\
\qquad\qquad\sum\limits_{\bu} \zeta (\bu ){\mathbf 1}(u_1=j)=y_j\;\forall\;j\in\cX\;
\hbox{ and}\,\,
\sum\limits_{\bv}\zeta (\bv )\vphi (\bv )\geq\eta ,\\
\qquad\qquad -{\diy\frac{1}{k}}\sum\limits_{\bv,\bv}\zeta (\bv )\log\,\tp^{\rm{so}}_{\bv,\bv'}
\leq h+\eps\bigg\}.\ena\ee
\ethm

For instance, take $k=2$ (i.e., the source process is a DTMC of order two,
and we work with $\vphi (i,j)$, $i,j\in\cX$). Then
$$\beal
B_{\ell ,2} (\eps,\eta )=\bigg\{y=(y_i)\in{\bbS}_\ell:\hbox{$\exists$ a vector}\;
z =(z_{ij})\in\bbR^{\ell^2}_+ \quad\hbox{such that}\quad\sum\limits_{j} z_{ij}=y_i,\quad\hbox{and}\\
\qquad\qquad\qquad\qquad\sum\limits_{i,j} z_{ij}\vphi (i,j)\geq\eta, \quad{\diy\frac{1}{2}}\sum\limits_{i,j,k,l}
z_{ij}\log\,p^{\rm{so}}_{ij,kl}\leq h+\eps\bigg\}.\ena$$

\br {\rm The bulk of the above analysis does not rely upon the
particular form of the two-digit WF $(i,j)\in\cX\times\cX\mapsto -
\log\,\tp^{\rm{st}}_{ij}$ related to the information rate of a string. The choice
of this WF (and of the upper bound $\sum T^{(n)}_{ij}\log \tp^{\rm{so}}_{ij}\leq h+\eps$
in \eqref{eq:setBn0}) was made in order to connect with the
Shannon NCT. In fact, the results stand up for any choice of a function
$(i,j)\mapsto\vphi_2(i,j)$. However, selecting $\cB_n$ with
${\diy\frac{1}{n}}\log p^{\rm{so}}_n(\cB_n)\leq\sigma <0$ would lead to a further
reduction of the memory volume needed to store set $\cB_n$.}
\er

\subsection{Examples}
{\bf A.} Let $\cX=\{0,1\}$ with $\ell =2$ (a binary alphabet).
Assuming that distributions $p^{\rm{st}}_n$ are generated by a DTMC, write the TM
$\tP^{\rm{st}}$ in the form
$$\tP^{\rm{st}}=\begin{pmatrix} 1-\alpha&\alpha\\ \beta&1-\beta\end{pmatrix}\,,
\quad\hbox{ with }\pi^{\rm{st}}_0=\frac{\beta}{\alpha +\beta},\,\,\,
\pi^{\rm{st}}_1=\frac{\alpha}{\alpha +\beta},$$
where $\alpha,\beta\in (0,1)$.

The analysis of maximization in \eqref{eq:M*f1Pi*g0} for $\ell =2$ and given
$y =(y_0,y_1)\in\bbS_2$, with $0<y_0,y_1<1$, can be done in a straightforward
(although tedious) manner. Recall: we want to find the maximum, in $0<\tu<1$,
of the expression
\be\label{eq:maximi}\beal
\diy y_0\ln\frac{\tu}{(1-\alpha)\tu + \alpha(1-\tu)} + y_1\ln\frac{1-\tu}{\beta\tu + (1-\beta )(1-\tu)}\\
\qquad\qquad\qquad\qquad \diy =-y_0 \ln(1-\alpha + \alpha\tw) - y_1\ln\left(1-\beta
+\frac{\beta}{\tw}\right),\ena\ee
with $\tw=\diy\frac{1-\tu}{\tu}\in (0,\infty )$. It is convenient to maximize in $\tw$.
To this end, we solve
$$0=\frac{\partial}{\partial\tw}\left[-y_0 \ln(1-\alpha + \alpha\tw) - y_1\ln\left(1-\beta
+\frac{\beta}{\tw}\right)\right]$$
which is equivalent to the quadratic equation
$$y_0\alpha(1-\beta)\tw^2 + \alpha\beta(y_0-y_1)\tw - y_1\beta(1-\alpha) = 0.$$
A solution $\tw =K(y)$ has been identified in \cite{DM}:
\be\label{eq:Ktwost}\beal K(y)=\diy{\frac{1}{2\alpha(1-\beta)y_0}}\\
\qquad\times\bigg[-\alpha\beta(y_0-y_1) +\sqrt{(\alpha\beta(y_0-y_1))^2
+4\alpha\beta (1-\alpha)(1-\beta)y_0y_1}\;\bigg].\ena\ee

Then
\be\label{eq:M*twost}\rM^*(y)=-y_0\log\,(1-\alpha +\alpha K)-y_1\log\,(1-\beta +
\beta /K)\ee
and
$$\rM^*(y )=\begin{cases}-\log\;(1-\beta ),
&y_1=1,\\ -\log\;(1-\alpha ),&y_0=1.\end{cases}$$
It is true that $\rM^*(y )=0$ if and only if $y_0=\pi^{\rm{st}}_0$,
$y_1=\pi^{\rm{st}}_1$. Examples of graphs of function $y_0\in (0,1)\mapsto\rM^*(y)$
are given below. See also animations enclosed.

\vskip 3 truemm

\noindent
\includegraphics[scale=0.7]{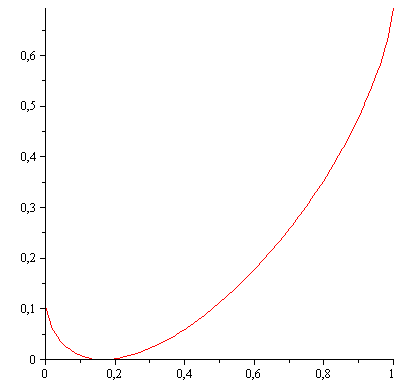}\includegraphics[scale=0.7]{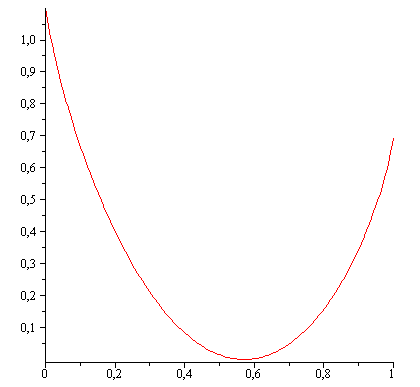}\\
\cl{$\rM^*(\alpha= 1/2, \beta= 1/10, y_0)$  \qquad\qquad\qquad\quad $\rM^*(\alpha= 1/2, \beta= 2/3, y_0)$}

\vskip 4 truemm

Accordingly, for $z =(z_{00},z_{01},z_{10},z_{11})\in\bbS_{4}$
with $z_{ij}\geq 0$ and $\sum\,z_{ij}=1$\,, the value $\Pi^*(z)$ is given as
follows. Set: $z^*=z_{00}+z_{01}$, $1-z^*=z_{10}+z_{11}$ and $y^*=(z,1-z)\in\bbS_2$.
Then
\be\label{eq:Pi*twost}\Pi^*(z)=\rM^*(y^*).\ee

\medskip

{\bf B.} Still with $\cX=\{0,1\}$ take $p^{\rm{st}}_n=p^{\rm{eq}}_n$. Suppose
the source distributions $p^{\rm{so}}_n$ are generated by a DTMC
with a transition matrix
$\tP^{\rm{so}}=\begin{pmatrix}\tp^{\rm{so}}_{00}&\tp^{\rm{so}}_{01}\\
\tp^{\rm{so}}_{10}&\tp^{\rm{so}}_{11}\end{pmatrix}$, with $\diy
\pi^{\rm{so}}_0=\frac{\tp^{\rm{so}}_{10}}{\tp^{\rm{so}}_{01}+\tp^{\rm{so}}_{10}},
\;\ \pi^{\rm{so}}_1=
\frac{\tp^{\rm{so}}_{01}}{\tp^{\rm{so}}_{01}+\tp^{\rm{so}}_{10}}$,
the value $\gam (\tP^{\rm{so}},\eps,\eta )$ from \eqref{eq:kapeq0} equals
\be\label{eq:kap2} \gam (\tP^{\rm{so}},\eps,\eta )=\sup\;\Big[H ( y):\;
y=(y_0,y_1)\in\bbS_2,\,y_1\in A_2 \Big]\ee
where interval $A_2 = A_2 (\tP^{\rm{so}},\eps,\eta )\subseteq [0,1]$
is given by
\be\label{eq:intA2}\beal A_2 =\Big\{0\leq u\leq 1:\;(1-u)\vphi (0)+u\vphi (1)
\geq\eta,\;\;\exists \,\,z^{(j)}=(z_{j1},z_{j2})\in\bbR^2_+, \,j=0,1,\\
\qquad\hbox{ such that }
z_{00}+z_{01}=1-u, \,\,z_{10}+z_{11}=u,\,\,
-\sum\limits_{i,j=0}^1 z_{ij}\log\,\tp^{\rm{so}}_{ij}\leq h+\eps\Big\}.\ena\ee
Here $H(y)=-y_0\log\,y_0-y_1\log\,y_1$ and $h=-\sum\limits_{i,j=0}^1
\pi^{\rm{so}}_i\tp^{\rm{so}}_{ij}\log p^{\rm{so}}_{ij}$.

Further, assuming $\tp^{\rm{so}}_{12}+\tp^{\rm{so}}_{21} =1$, the above matrix
$\tP^{\rm{so}}$  has a repeated row $\mbp =(1-\tp,\tp )$ where $0<\tp<1$.
It yields an IID source, and the formula \eqref{eq:kap2} for $\gam$ simplifies.
We write $h=-(1-\tp )\log\,(1-\tp )-\tp\log\,\tp$, and
\be\label{eq:kap3} \gam (\tp;\eps ,\eta ) =
\sup\;\Big[H ( y):\;
y=(y_0,y_1)\in\bbS_2,\,y_1\in D_2 \Big].\ee
Here  interval $D_2 = D_2 (\tP^{\rm{so}},\eps,\eta )\subseteq [0,1]$
is given by
\be\label{eq:intD2}\beal D_2=\Big\{0\leq u\leq 1:(1-u)\vphi (0)+y\vphi (1)
\geq\eta,\\
\qquad\qquad-(1-u)\log\,(1-\tp)-u\log\,\tp\leq h+\eps\Big\}\,.\ena\ee
We reiterate: the maxima in \eqref{eq:kap2} and \eqref{eq:kap3} are
attained either at
$u=1/2$ -- when $1/2\in D_2(\eps ,\eta )$, or at the nearest endpoint.

\subsection{}
Next, we are going to (quickly) discuss multiplicative WFs  $\phi_n(\bx)
=\prod\limits_{i=0}^{n-1}\psi (x_i)$.
Assuming that function $\psi$ is strictly positive, consider selecting strings with
$\phi_n(\bx_0^{n-1})\geq\ree^{n\eta}$. Passing to the logarithms yields

\bthm\label{Thm:2.3}
Under the assumptions of Theorem \ref{Thm:2.2}, select
strings  $\bx\in\cC^n$ where
\be\label{eq:setCn0} \cC_n=\Big\{\bx :\,\sum\limits_{i}
U^{(n)}_i\log\psi (i)\geq\eta,\,\, -\sum\limits_{i,j}
T^{(n)}_{ij}\log\tp^{\rm{so}}_{ij}\leq h+\eps\Big\}.\ee
Then, with $c_n=\#\cC_n$,
\be\label{eq:cPn}\lim\limits_{n\to\infty}\frac{1}{n}\log\,c_n=
\iota (\tP^{\rm{so}},\eps,\eta ).\ee
Here $\iota (\tP^{\rm{so}},\eps,\eta )$ is given by \eqref{eq:kapeq0}
with $\vphi$ replaced by $\log\psi$.
\ethm

Various generalizations can be achieved by following the same line of
argument as for additive WFs.

\section{The case of a general Markov source}\label{sect:contin}

When the alphabet set is large (or continuous), one can
use a general theory where the source output is represented by a
sequence of
points in a space $\cX$ with some structure. Such a situation
is typical when one stores analogous data. In particular, the volume
in $\cX$ may be represented by a given measure $\nu$ with
$\rV=\nu (\cX )<\infty$. As above, the volume in $\cX^n$ may be
associated with the product-measure $\nu^n$
or have a more involved form. Our aim here is similar: to assess the
amount of volume needed to store valuable strings
$\bx =(x_0,\ldots , x_{n-1})\in\cX^n$.
A normalized product-volume $\diy\frac{\nu^n}{\rV^n}$ yields a
probability measure, an analog of $p^{\rm{eq}}_n$;
asymptotic analysis of the volume is reduced to that of $p^{\rm{eq}}_n$.
More generally, as in Sect \ref{sect:finite}, we discuss the case where the volume is
of the form $\rV_np^{\rm{st}}_n(\bx)$, assuming that $p^{\rm{st}}_n$ as
well as the source distribution $p^{\rm{so}}_n$ are generated by
DTMCs with state space $\cX$. Transition matrices $\tP^{\rm{st}}$
and $\tP^{\rm{so}}$ are replaced with
transition functions $\Big\{\tP^{\rm{st}}(x,A)\Big\}$ and
$\Big\{\tP^{\rm{so}}(x,A)\Big\}$, $x\in\cX$,
$A\subseteq\cX$; standard measurability conditions apply by default.

\subsection{}
From now on  $\cX$ is a
Polish space with a chosen metric,
$\rC_{\rm b}(\cX)$ is the space of continuous bounded (real)
functions $f:\cX\to\bbR$ with the sup-norm
and $\sP(\cX)$ the space of Radon probability measures $\ups$ on
$\cX$ with the L\'evy--Prokhorov metric.
(Then  $\sP(\cX)$ is a Polish space.) In a similar manner, consider
the space $\sP(\cX\times\cX)$.
Let us  fix a non-negative finite Radon measure $\nu$ on $\cX$ and
designate $\sP_\nu =\sP_\nu(\cX)$ to be the set of measures absolutely
continuous relative to $\nu$. Next, let
$\sP_{\nu\times\nu}=\sP_{\nu\times\nu}(\cX\times\cX)$ designate the
set of measures absolutely continuous relative to $\nu\times\nu$.

We suppose that measures $\tP^\bullet (x,\,\cdot\,)$ are absolutely
continuous relative to $\nu$ and work with the corresponding transition
densities
$\tp^\bullet (x,x')=\diy\frac{\tP^\bullet (x,\rd x')}{\nu (\rd x')}$,
$x,x'\in\cX$. For $\tp^{\rm{st}} (x,x')$ we also adopt Assumption (U)
from \cite{DZ}, Ch. 6.3. (There exists a host of weaker conditions;
see Assumptions (H-1), (H-2) from \cite{DZ}, Ch. 6.3 and
from \cite{DS}, Ch. 5.4, leading to more involved formulas.)
For $\tp^{\rm{so}} (x,x')$ we assume ergodicity under a unique
equilibrium distributions $\pi^{\rm{so}}\in\sP_\nu$ and suppose that
the integral giving the IER converge absolutely:
\be\label{eq:entrora1}h=h^{\rm{so}}=\int_{\cX\times\cX}
\log\,\tp^{\rm{so}}(x,x')\pi^{\rm{so}} (\rd x)\,\tP^{\rm{so}}
(x,\rd x').\ee
As was said, $p^{\rm{st}}_n$ stands for the probability measure
on $\cX^n$ generated by the DTMC with transition density
$\tp^{\rm{st}}(x,y)$ under a given initial distribution $\lam$.

Let $\bx\in\cX^n$. Following \eqref{eq:UT}, consider empirical
measures $\bU^{(n)}=\bU^{(n)}(\bx )$ $\in\sP (\cX)$ and $\bT^{(n)}
=\bT^{(n)}(\bx )\in\sP (\cX\times\cX)$:
\be\label{eq:UTg}\bU^{(n)}=\frac{1}{n}\;
\sum\limits_{j=0}^{n-1}\delta_{x_j},\qquad
\bT^{(n)}=\frac{1}{n-1}\;\sum\limits_{j=0}^{n-2}\delta_{x_j,x_{j+1}}. \ee
Here $\delta$ stands for the Dirac mass.

According to standard LDP results,  $\bU^{(n)}$ and $\bT^{(n)}$ satisfy
the full LDP (in $\sP_\nu \times\sP_{\nu\times\nu}$)
with good convex LDR functions $\rM^*(\ups )$ and $\Pi^*(\tau )$,
$\ups\in\sP_\nu$, $\tau\in\sP_{\nu\times\nu}$.
See \cite{DZ}, Ch 6.3, particularly, Theorem 6.3.8.
Moreover, $\rM^*$ and $\Pi^*$ can be specified as follows.

\begin{itemize}
  \item (i) When  $\ups\in\sP(\cX)\setminus\sP_\nu $ or
$\tau\in\sP(\cX\times\cX)\setminus\sP_{\nu\times\nu} $, we have  $\rM^*(\ups )
=\Pi^*(\tau )=\infty$.
  \item (ii) 
For $\ups\in\sP_\nu$ and $\tau\in\sP_{\nu\times\nu}$,
\be\label{eq:M*f1Pi*g1}\beac\rM^*(\ups )=
\sup\,\Big[\int_\cX\log\,{\diy\frac{m(x)}{\tP^{\rm{st}}m\,(x)}}\,
\ups (\rd x):\,m\in\rC_\rb (\cX),\,m\geq 1\Big],\\
\Pi^*(\tau )=\sup
\Big[\int_\cX\log\,{\diy\frac{m(x)}{\tP^{\rm{st}}m\,(x)}}\,
\tau (\cX\times \rd x):\,m\in\rC_\rb (\cX),\,m\geq 1\Big]\ena\ee
where $\tP^{\rm{st}}m\,(x)=\int_\cX\tp^{\rm{st}}(x,y)m(y)\nu (\rd y)$.
\end{itemize}
Cf. \cite{DS}, Ch. 4.1, and \cite{DZ}, Ch. 6.5 (detailed references
have been given at the beginning of Sect 2.2).

\subsection{}
For an additive WF $\phi_n(\bx_0^{n-
1})=\sum\limits_{j=0}^{n-1}
\vphi (x_j)$ we assume that function $\vphi :\cX\to\bbR$ is continuous.
We want to select strings from $\cB_n=\cB_n(\eps ,\eta)$ where
\be\label{eq:cBng}\beal\cB_n=\bigg\{\bx\in\cX^n:\,\,
{\diy\frac{1}{n}}\sum\limits_{i=0}^{n-1}\vphi (x_i)\geq
\eta\,\,\hbox{ and}\,\, -{\diy\frac{1}{n-1}}\sum\limits_{i=0}^{n-2}
\log\,\tp^{\rm{so}}(x_{i},x_{i+1})\leq h+\eps\bigg\} \ena \ee
and $h$ is given in \eqref{eq:entrora1}. Equivalently,
\be\beal
\cB_n=\bigg\{\bx:\,\,\int_\cX\vphi (x)\bU^{(n)}(\rd x)\geq \eta\,\,
\hbox{ and}\\
\qquad -\int_{\cX\times\cX}
\log\,\tp^{\rm{so}}(x,x')\bT^{(n)}(\rd x\times\rd x')\leq h +
\eps\bigg\}.\ena \ee

Now, with $\Pi^*(\tau )=\Pi^*(\tP^{\rm{st}},\tau )$ as in
\eqref{eq:M*f1Pi*g1}, set:
\be\label{eq:kap11}
\kap (\eps, \eta )=\kap (\eps,\eta , \tP^{\rm{st}},\tP^{\rm{so}})=-\inf\;\Big[\Pi^*(\tau ):\;
\tau\in B\Big],\ee
where set $B=B(\tP^{\rm{so}},\eps ,\eta )\subset\sP_{\nu\times\nu}$
is given by
\be\label{eq:setBg}\beal B=
\bigg\{\tau :\;\int_{\cX\times\cX}\vphi (x')\tau
(\rd x\times\rd x')\geq\eta\;\hbox{ and}\\
\qquad\quad -\int_{\cX\times\cX} \log\,\tp^{\rm{so}}(x,x')\tau
(\rd x\times\rd x')\leq h+\eps\bigg\}.\ena\ee

\bthm\label{Thm:3.1}
Under the above assumptions,
for all $\eps, \eta >0$, the following relation holds true:
\be\label{eq:consLDP}\kap (\eps, \eta )=\lim\limits_{n\to\infty}\,
{\diy\frac{1}{n}}\,
\log p^{\rm{st}}_n\big(\cB_n).\ee
\ethm

In the case of volume $\nu^n$ in $\cX^n$, the above formulas simplify.
Let us set:
\be\label{eq:gameqg}\gam (\eps,\eta )
=\gam (\tP^{\rm{so}},\eps,\eta )=\inf\;\Big[H (\ups):\;
\ups \in A\Big].\ee
Here set $A=A(\tP^{\rm{so}},\eps,\eta )\subset\sP_\nu$ is given by
\be\label{eq:Aeqg}\beal A
 =\bigg\{\ups :\;\int_\cX
\vphi (x)\ups (\rd x)\geq\eta\;\hbox{ and $\;\exists\;$ a measure}
\, \tau\in\sP_{\nu\times\nu}\,
\hbox{ with}\\
\quad\tau (\cX\times\rd x')=\ups (\rd x')\;
\hbox{ and }-\int_{\cX\times\cX}\log\,\tp^{\rm{so}}
(x,x')\tau (\rd x\times\rd x')\leq h+\eps\bigg\}\ena\ee
and for $\mu\in\sP_\nu$ with $m(x)=\diy\frac{\mu (\rd x)}
{\nu (\rd x)}$,
\be\label{eq:Entro1}H(\mu )=-\int_\cX m(x)\log m(x)\nu (\rd x). \ee

If $\tp^{\rm{so}}(x,x')=\tp^{\rm{so}}(x')$ (i.e., in the case of
IID source outputs), the formula for $\gam$ is further streamlined. It
is expressed in terms of density $p^{\rm{so}}(x')$: here the entropy
rate $h=-\int\limits_{\cX}\tp^{\rm{so}}(x')\log\,\tp^{\rm{so}}(x')\nu
(\rd x')$, and 
\be\label{eq:gamIIDg} \gam (\eps,\eta )=
\inf\;\Big[H (\ups):\;\ups \in D\Big],\ee
with $D=D (\tP^{\rm{so}},\eps,\eta )\subset\sP_\nu$:
\be\label{eq:setDg}\beal D=\Big\{\ups :\;\int_{\cX}
\vphi (x)\ups (\rd x)\geq\eta,\,\, -\int_{\cX}\log\,\tp^{\rm{so}}(x)\ups (\rd x)\leq h+\eps\Big\}.
\ena\ee

The above construction leads to following result.

\bthm\label{Thm:3.2} Let the source process be an ergodic DTMC
with states $x,x'\in\cX$, transition densities $\tP^{\rm{so}}
=\{\tp^{\rm{so}}(x,x')\}$ and equilibrium
density $\pi^{\rm{so}}(x)$. Let $h$ stand for the IER.
Consider the volume $v_n=\nu^n(\cB_n)$ of set $\cB_n\subset\cX^n$
as in \eqref{eq:cBng}. Then
\be\label{eq:cBn1}\lim\limits_{n\to\infty}\frac{1}{n}\log\,v_n=
\gam (\tP^{\rm{so}},\eps,\eta )\ee
where $\gam (\eps,\eta )$ is given by \eqref{eq:gameqg}.
For an IID source, with $\tp^{\rm{so}}(x,x')=\tp^{\rm{so}}(x')$, one uses
Eqn \eqref{eq:gamIIDg}.
\ethm

\br {\rm The entropy functional $\mu\in\sP_\nu\mapsto H(\mu )$ in
\eqref{eq:Entro1} is concave, and set $D$ in \eqref{eq:gamIIDg} is convex.
It is tempting to conjecture that if $D$ does not contain the probability measure
$\nu/\rV$ (the global maximizer of $H$ in $\sP_\nu$) then the maximum
of $H(\ups )$ over $D$ is attained at a unique point lying in a (suitably defined)
boundary $\partial D$. This direction needs further exploring; many aspects of
convexity and related topics of optimisation are discussed in \cite{AB}--\cite{BV}.}
\er

\subsection{}
For a multiplicative WFs  $\phi_n(\bx)=\prod\limits_{i=0}^{n-1}\psi (x_i)$ with
strictly positive one-digit factor $\psi (x)$, we obtain the following assertion:

\bthm\label{Thm:3.3} Under the assumptions of Theorem \ref{Thm:2.2}, select
strings  $\bx\in\cC^n$ where $\cC_n\subset\cX^n$ is as in \eqref{eq:setCn0}.
Then, for $w_n=\nu^n(\cC_n)$,
\be\label{eq:cPng}\lim\limits_{n\to\infty}\frac{1}{n}\log\,w_n=
\iota (\tP^{\rm{so}},\eps,\eta ).\ee
Here $\iota (\tP^{\rm{so}},\eps,\eta )$ is given by \eqref{eq:gameqg} and
\eqref{eq:Aeqg} with $\vphi$ replaced by $\log\psi$.
\ethm

\section*{Concluding remarks}
The paper discusses the problem of storing
`valuable' data (digital or analogous) selected on the basis of the rate of
a utility/weight function.
The storage space is treated as an expensive commodity that should be
provided and organized in an efficient manner.
The issue of reducing and organising storage space is addressed from
a probabilistic point of view which is an extension of the Shannon data-compression
principle (the Shannon Noiseless coding theorem). More precisely, the storage volume
is assessed via the theory of large deviations. The emerging optimization problem
is highlighted and explained through examples.

\medskip
\noindent
{\bf Acknowledgement} YS thanks the Math. Department, Penn State University, for hospitality
and support. IS thanks the Math. Department, University of Denver, for support and hospitality.


\begin{thebibliography}{10}

\bibitem{CT} T. Cover, J. Thomas. {\it Elements of Information Theory.} New York: Wiley, 2006.

\bibitem{KS} M. Kelbert and Y. Suhov. {\it Information Theory and Coding by
Example.} Cambridge: CUP, 2013.

\bibitem{SSYK} Y. Suhov, I. Stuhl, S. Yasaei Sekeh, M. Kelbert. Basic
inequalities for weighted entropies. {\it Aequatioines Math.}, (2016); DOI: 10.1007/s00010-015-0396-5.

\bibitem{SYS} Y. Suhov, I. Stuhl, S. Yasaei Sekeh. Weighted Gaussian
entropy and determinant inequalities. arXiv:1505.01753.

\bibitem{SSK} Y. Suhov, I. Stuhl, M. Kelbert. Weight functions and log-optimal 
investment portfolios. arXiv:1505.01437.

\bibitem{SS1} Y. Suhov, I. Stuhl. Weighted entropy rates. (In progress).

\bibitem{AC} P. Algoet, T. Cover. A sandwich proof of the
Shannon--McMillan--Breiman theorem. {\it Ann. Probab.}, {\bf 16}, No. 2
(1988), 899--909.


\bibitem{DM} K. Duffy,  A.P. Metcalfe. The large deviations of estimating
rate functions. {\it J. Appl. Prob.}, {\bf 42} (2005), 267-274.




\bibitem{Bu} A. Bucklew. {\it Large deviation techniques in decision, simulation,
and estimation.} New York, NY: Wiley, 1990.

\bibitem{DZ} A. Dembo, O. Zeitouni. {\it Large deviations techniques and applications.}
2nd Edition/corrected printing. Berlin: Springer, 2010.

\bibitem{DS} J.-D. Deuschel, D.W. Stroock. {\it Large deviations.} San Diego, CA:
Academic, 1989; reprint: Providence, R.I.: AMS Chelsea Publishing, 2000.

\bibitem{DE} P. Dupuis, R.S. Ellis. {\it A weak convergence approach to the theory of large
deviations.}  New York, NY: John Wiley \& Sons, 1997.

\bibitem{El} R.S. Ellis. {\it Entropy, large deviations, and statistical mechanics.} New York, NY:
Springer, 1985; reprint: 2006.

\bibitem{FK} J. Feng, T.G. Kurtz, {\it Large deviations for stochastic processes.}
Providence, RI.:  AMS, 2006.

\bibitem{Ho} F. den Hollander. {\it Large deviations.} Fields Institute Monograph {\bf 14}.
Providence, RI: AMS, 2000.

\bibitem{Pu} A. Puhalskii. {\it Large deviations and idempotent probability.}
Boca Raton {\it et al.}: Chapman \& Hall/CRC, 2001.



\bibitem{SW} A. Shwartz, A. Weiss, {\it Large deviations for performance
analysis.} Stochastic Modeling Series. London: Chapman and Hall, 1995.


\bibitem{St} D. Stroock. {\it An introduction to the theory of large
deviations.} New York, NY: Springer, 1984.

\bibitem{V1} S. R. S. Varadhan. {\it Large deviations and applications.}
Philadelphia, PA: SIAM, 1984.

\bibitem{V2} S. R. S. Varadhan. Large deviations. {\it Ann. Prob.}, {\bf 36}
(2008), 397--419.


\bibitem{AB} C.D. Aliprantis, K.C. Border. {\it Infinite dimensional analysis.
A Hitchhiker's guide},  3rd Ed. Berlin: Springer, 2006.

\bibitem{Be} G.A. Beer. Topologies on closed and closed convex sets.
Dordrecht: Kluwer, 1993.

\bibitem{BV} S. Boyd, L. Vanderberghe. {\it Convex optimization.} Cambridge:
CUP, 2004.

\end{thebibliography}
\end{document}